 \documentclass[reprint, aps, prapplied,preprintnumbers, amsmath,amssymb,floatfix, longbibliography]{revtex4-2}

\usepackage{graphicx}
\usepackage{dcolumn}
\usepackage{multirow}
\usepackage{bm}

\usepackage{comment}  

\usepackage{xcolor} 
\usepackage{ulem} %

\usepackage{tikz, tkz-euclide, ifthen}
\definecolor{gold}{rgb}{0.95, 0.69, 0.24}
\definecolor{grey}{rgb}{0.57, 0.57, 0.57}

\pgfset{
	foreach/parallel foreach/.style args={#1in#2via#3}{evaluate=#3 as #1 using {{#2}[#3-1]}},
}

\definecolor{amethyst}{rgb}{0.6, 0.4, 0.8}

\definecolor{colorblue}{RGB}{4,4,236}

\usepackage{hyperref}

\definecolor{colorblue}{RGB}{4,4,236}
\hypersetup{
    colorlinks=true,  
    linkcolor=colorblue,   
    citecolor=colorblue,    
    urlcolor=colorblue, 
    filecolor=colorblue   
}

\begin{document}

\title{Volume-Preserving Deformation of Honeycomb Wire Media Enables Broad Plasma Frequency Tunability}



\author{
Denis Sakhno$^{1,*}$, Jim A. Enriquez$^{1,2,*}$, Pavel A. Belov$^{1,3}$\\[1ex]
\textit{$^{1}$School of Physics and Engineering, ITMO University, Kronverksky Pr. 49, 197101, St. Petersburg, Russia}\\
\textit{$^{2}$Physics Department, National University of Colombia, 111321, Bogota, Colombia}\\
\textit{$^{3}$School of Engineering, New Uzbekistan University, Movarounnahr str. 1, 100000, Tashkent, Uzbekistan}\\[1ex]
\small Emails: denis.sakhno@metalab.ifmo.ru, jim.enriquez@metalab.ifmo.ru\\
\small $^{*}$These authors contributed equally to this work
}



\date{\today}

\begin{abstract}

We demonstrate significant tunability of the plasma frequency in a wire medium by mechanically deforming a lattice of parallel metallic wires arranged at the nodes of a honeycomb structure. Numerical simulations predict up to 78\% tunability and a proof-of-concept experiment confirms 64\%, surpassing previously reported values for tunable wire media.

\end{abstract}

\maketitle

Metamaterials are artificially engineered structures designed to possess properties not commonly observed in nature \cite{engheta2006metamaterials,kadic20193d, sakoda2019electromagnetic, alu2024metamaterials, Alu2025}. One notable class of metamaterials, known as wire media, consists of periodically arranged conductive wires embedded in a host medium or in free space \cite{simovski2012wire}. Wire media are characterized by pronounced spatial dispersion, even at low frequencies \cite{belov2003strong, simovski2004}, enabling advanced manipulations of electromagnetic fields \cite{simovski2012wire, belov2006resolution, Palikaras2010, fernandes2012cherenkov, silveirinha2012radiation}.

A wire medium \cite{Brown, Rotman, pendry1998low} refers to a structure composed of parallel metallic wires periodically arranged in a perpendicular plane. This wire metamaterial prevents wave propagation below a characteristic frequency known as the plasma frequency. This frequency depends on the periodicity of the material, i.e. on the geometry of the unit cell \cite{belov2002dispersion}, on the number of wires and their arrangement within the unit cell \cite{fernandes2013fano, kowitt2023tunable}. The plasma frequency determines the dispersion of the wire medium~\cite{belov2002dispersion, simovski2004} and sets the epsilon-near-zero condition \cite{Silveirinha2006, Alu2025}, which can be tailored to enhance antenna performance~\cite{Jafargholi2024, Boas2025}. 

\begin{figure}[h]
        \begin{minipage}{1.00\linewidth}
		\center{ 
			\pgfset{
	foreach/parallel foreach/.style args={#1in#2via#3}{evaluate=#3 as #1 using {{#2}[#3-1]}},
}
\definecolor{gold}{rgb}{0.95, 0.69, 0.24}
\definecolor{grey}{rgb}{0.57, 0.57, 0.57}

\begin{tikzpicture}[scale=1.04, transform shape]
    \def\radii{1/3}; 
    \def\period{2.5};
    \def\rc{0.05*\period}
    
    \coordinate(A0) at (0, 0); 
    \coordinate(A1) at (\period, 0); 
    \coordinate(A2) at (0.5 * \period, {0.5 * sqrt(3) * \period}); 
    \coordinate(A3) at (1.5 * \period, {0.5 * sqrt(3) * \period});
    \coordinate(A4) at (2 * \period, 0);
    \coordinate(A5) at (-0.5 * \period, {0.5 * sqrt(3) * \period});
    \coordinate(A6) at (2.5 * \period, {0.5 * sqrt(3) * \period});
    
    \def \allcenters  {A0, A1, A2, A3, A4, A5, A6}
    \def \centers {A0, A1, A2, A3, A4}
    \def\cylcol{gold}
    
    \def\ly{0};
    \def\lyy{0.5 * sqrt(3) * \period};
    \def\cy{(\ly + \lyy) / 2}
    \def\d{0.1}
    
    \draw[-, thick, white](0, {\cy-\period*(1+\d)}) -- (0, {\cy+\period*(1+\d)});
    \draw[-, thick, white]({-\period/2}, \ly) -- ({2*\period}, \ly);
    \draw[-, thick, white]({-\period*0.5}, {\lyy}) -- ({\period*2}, {\lyy});
    \draw[-, thick, white]({-\period/2}, {\cy}) -- ({2*\period}, {\cy});

    \def \periodMarkShift {\period * 0.65}
    \def \marginPeriod {0.1 * \period}

    \draw[-, line width=0.25pt, gray]({-0.5*\period}, {0}) -- ({-0.5*\period}, {-\periodMarkShift - \marginPeriod});
    \draw[-, line width=0.25pt, gray]({0.5*\period}, {0}) -- ({0.5*\period}, {-\periodMarkShift - \marginPeriod});
    
    \draw[<->, line width=0.75pt, black]({-0.5*\period}, {-\periodMarkShift}) -- ({0.5*\period}, {-\periodMarkShift}) node[midway, below, yshift=0pt] {\large $a$};
    
    \def \margin {0.1}
    \draw[white] ({-0.5 * \period - \period * \margin}, {0}) -- ({2.5 * \period + \period * \margin}, {0});
    
    \draw[->, line width=0.5pt, grey] ({0}, {0}) -- ({0.9*\period}, 0) node[left, xshift=3pt, yshift=-7pt] {$x$};
    \draw[->, line width=0.5pt, grey] (0, 0) -- (0, {{\period * 3 / 4}}) node[left, , xshift=-1pt, yshift=-3pt] {$y$};
    
    \foreach \center in \centers {
        \filldraw [gray] (\center) circle (0.5pt); 
    }
    
    \foreach \center in \centers {
        \foreach \i in {0, 1, 2, 3, 4, 5} {
            \coordinate(this) at ({\period*\radii*cos(\i * 60)}, {\period*\radii*sin(\i * 60)});
            \filldraw [black, fill=\cylcol, line width=0.25pt] (\center){} + (this){} circle (\rc);
        }
    }
    
    \foreach \i in {0, 1, 5} {
        \coordinate(this) at ({\period*\radii*cos(\i * 60)}, {\period*\radii*sin(\i * 60)});
        \filldraw [black, fill=\cylcol, line width=0.25pt] (A5){} + (this){} circle (\rc);
    }
    \foreach \i in {2, 3, 4} {
        \coordinate(this) at ({\period*\radii*cos(\i * 60)}, {\period*\radii*sin(\i * 60)});
        \filldraw [black, fill=\cylcol, line width=0.25pt] (A6){} + (this){} circle (\rc);
    }
    
    \def\circX{\period*\radii*cos(4 * 60)};
    \def\circY{\period*\radii*sin(4 * 60)};
    \draw[->, black, line width=0.5pt]({\circX-3*\rc *cos(0)}, {\circY-3*\rc * sin(0)}) -- ({\circX-\rc*cos(0)}, {\circY-\rc*sin(0)});
    \draw[->, black, line width=0.5pt]({\circX+3*\rc *cos(0)}, {\circY+3*\rc * sin(0)}) -- ({\circX+\rc*cos(0)}, {\circY+\rc*sin(0)}) node[midway,xshift=7pt, yshift=-7pt] {$2r_0$};
    
    \def\ucsz{\period / 2 / cos(30)};
    \foreach \i in {0, 1, 2, 3, 4, 5} { 
        \coordinate(this) at ({\ucsz*cos(30+(\i)*60)}, {\ucsz*sin(30+(\i)*60)});
        \coordinate(prev) at ({\ucsz*cos(30+(\i-1)*60)}, {\ucsz*sin(30+(\i-1)*60)});;
        \draw[solid, line width=0.5pt, black](prev) -- (this);
    }
    
    \def\shiftX{\period * (1 + cos(60))};
    \def\shiftY{ \period * sin(60)};
    
    \coordinate(B0) at ({-0.75 * \period + \shiftX}, {-0.25 * sqrt(3) * \period + \shiftY});
    \coordinate(B1) at ({\period - 0.75 * \period + \shiftX}, {-0.25 * sqrt(3) * \period + \shiftY});
    \coordinate(B2) at ({\period * cos(60) - 0.75 * \period + \shiftX}, {\period * sin(60) - 0.25 * sqrt(3) * \period + \shiftY});
    \coordinate(B3) at ({0.75 * \period + \shiftX}, {0.25 * sqrt(3) * \period + \shiftY});
    
    
    
    \def \shiftLen {0.15 * \period}
    \def \lenTick {1.5}  
    
    \coordinate(Arrow1) at ({\period}, {-\shiftLen}); 
    \coordinate(A12) at ({\period}, {-\lenTick * \shiftLen}); 
    \coordinate(Arrow4) at ({2 * \period}, {-\shiftLen});
    \coordinate(A42) at ({2 * \period}, {-\lenTick * \shiftLen});


    \coordinate(Arrow12) at ({\period - \shiftLen * cos(30)}, {\shiftLen * sin(30)});
    \coordinate(A13) at ({\period - \lenTick * \shiftLen * cos(30)}, {\lenTick * \shiftLen * sin(30)});
    \coordinate(Arrow3) at ({1.5 * \period - \shiftLen * cos(30)}, {0.5 * sqrt(3) * \period + \shiftLen * sin(30)});
    \coordinate(A32) at ({1.5 * \period - \lenTick * \shiftLen * cos(30)}, {0.5 * sqrt(3) * \period + \lenTick * \shiftLen * sin(30)});

    

    \def \marginR {0.105 * \period}

    \draw[-, line width=0.25pt, gray]({0}, {0}) -- ({0}, { 1.5 * \marginR});
    \draw[-, line width=0.25pt, gray]({{-\period * \radii}}, {0}) -- ({-\period * \radii}, { 1.5 * \marginR});

    \filldraw [gray] ({-(\period * \radii)}, {0}) circle (0.5pt); 

    \draw[<->, line width=0.75pt, black]({0}, {\marginR}) -- ({-(\period * \radii)}, {\marginR}) node[midway, below, yshift=0pt] {\large $R$};

\end{tikzpicture}
		}
            \put(-241, 140){\large {(a)}}
	\end{minipage}
        \vfill
        \vspace{5pt}
	\begin{minipage}{1.00\linewidth}
		\center{
                \input{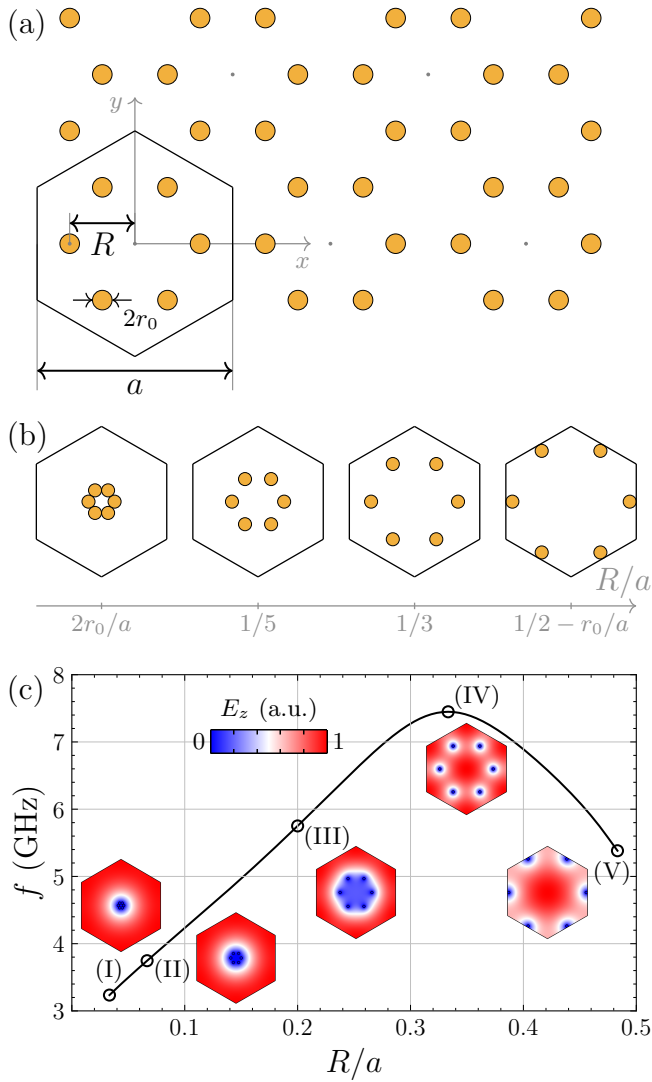}
		}
            \put(-247, 243){\large {(b)}}
            \put(-247, 147){\large {(c)}}
	\end{minipage}
	\caption{
        (a) Geometry of a wire metamaterial with a honeycomb lattice of period $a$. A hexagonal unit cell (solid black line) contains parallel wires of radius $r_0$. $R$ is the distance between the unit cell center and wires within a unit cell.
	(b) Breathing deformation of the structure visualization.
        (c) Numerical calculation of the metamaterial plasma frequency during the breathing deformation. Geometry parameter $R$ changes from $2r_0$ to $a/2-r_0$ ($a=3$ cm, $r_0=0.5$ mm) in steps of $10^{-3}a$ (451 data points)}. Insets in the plot show field distributions of the plasma mode for each highlighted $R/a$ value. 
	 \label{fig:deformation} 
\end{figure}

The interest in wire metamaterials has grown in recent years due to their potential applications in tunable resonators for dark matter searches~\cite{kowitt2023tunable, Lawson2019, balafendiev2022resonator}. In particular, the ALPHA consortium employs a wire-medium-filled cavity as a key component of its experimental setup for axion detection -- a leading dark matter candidate~\cite{millar2023alpha}. A key advantage of wire-medium-filled cavities over conventional designs is their ability to operate at a target resonance frequency largely independent of the cavity volume.
In this work, we study a wire metamaterial based on a hexagonal lattice with lattice constant $a$, as illustrated in Fig.~\ref{fig:deformation}(a). Each unit cell (the hexagonal region bounded by solid black lines in Fig.~\ref{fig:deformation}(a)) contains six metallic wires of radius $r_0$, symmetrically placed to form a regular hexagon inscribed in a circle of radius $R$.
The size of a hexagon formed by the wires within each unit cell can be adjusted while preserving the central point and orientation of the hexagon. Variations in the geometric parameter $R$ (representing expansion and shrinkage of the hexagon of wires) are associated with what can be referred to as \textit{a breathing deformation} of the lattice (see Fig.~\ref{fig:deformation}(b)). This mechanical deformation preserves an overall volume of the metamaterial, making it a good candidate for applications in a tunable resonator for a dark matter search.

Alongside volume preservation, another crucial property for a potential tunable resonator filled with a wire medium is tunability of the working frequency -- in this case, the plasma frequency \cite{millar2023alpha}. The tunability in wire media formed by rectangular lattices has been previously investigated in Refs.~\cite{belov2002dispersion, kowitt2023tunable, sakhno2024anisotropy} while other lattices remain
unstudied in that context.

To demonstrate a tunability provided by the breathing deformation we performed a numerical simulation in COMSOL \cite{comsol} for an infinite wire medium (applying periodic boundary conditions to a hexagonal unit cell and calculating the lowest eigenmode in the $\Gamma$-point) with a period $a=3$ cm composed of infinite and perfectly electric conducting (PEC) wires of radii $r_0=0.5$ mm. The plot in Fig. \ref{fig:deformation}(c) shows a dependence of the plasma frequency on a deformation parameter $R$ changing from the minimal possible size equal to $2r_0$, where wires touch near the unit cell center, to the maximal $a/2-r_0$, where wires from neighboring cells touch near the unit cell boundary, with a step equal to $10^{-3}a$. 

Figure \ref{fig:deformation}(c) shows that the minimum plasma frequency $f_\text{min}\approx 3.236$ GHz is achieved when all six wires are grouped near the center of the unit cell ($R/a=1/30$). Conversely, when the wires are evenly spaced, such that the distance from any wire to its nearest neighbor is equal to $R=a/3$ ($R/a=1/3$), the metamaterial exhibits the maximum plasma frequency $f_\text{max}\approx 7.448$ GHz. In Table \ref{tab:tunability} plasma frequency values are provided for the points marked in Fig. \ref{fig:deformation}(c).

\begin{table}[h!]
  \centering
  \begin{tabular}{|c|c|c|}
    \hline
    Point & $R/a$ & $f$ (GHz)\\
    \hline
    I & $1/30$ & 3.236 
    \\ 
    II & $1/15$ & 3.747 
    \\ 
    III & $1/5$ & 5.751 
    \\ 
    IV & $1/3$ & 7.448 
    \\ 
    V & $29/60$ & 5.382 
    \\
    \hline
  \end{tabular}
  \caption{
  Plasma frequencies obtained from numerical simulations of infinite periodic structure ($a=3$ cm, $r_0=0.5$ mm), see Fig. \ref{fig:deformation} (c).
  }
  \label{tab:tunability}
\end{table}

Thus, our numerical simulations for the infinite structure formed by a hexagonal lattice demonstrate remarkable tunability in the wire medium
\begin{equation}
    \eta_\text{max} = 
    \frac{2(f_\text{max} - f_\text{min})}{f_\text{max} + f_\text{min}}
    \times 100 \% \approx 78.85\%,
    \nonumber
\end{equation}
which significantly surpasses the $\sim16\%$~\cite{kowitt2023tunable} and $\sim26\%$~\cite{balafendiev2025tunableepsilonnearzero} tunability reported in previous studies.



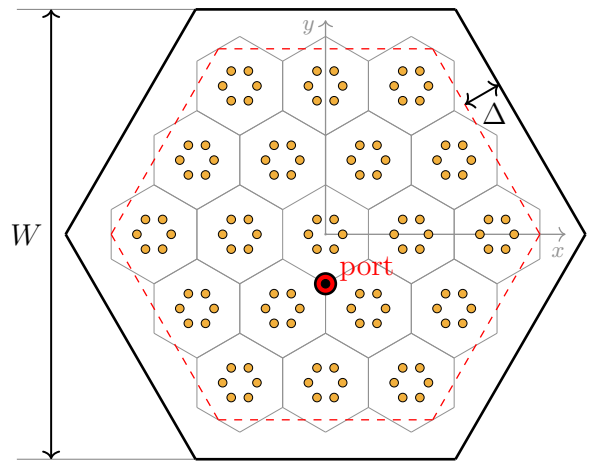
\begin{figure}[h]
	\begin{minipage}{1.00\linewidth}
		\center{
            \pgfset{
	foreach/parallel foreach/.style args={#1in#2via#3}{evaluate=#3 as #1 using {{#2}[#3-1]}},
}
\definecolor{gold}{rgb}{0.95, 0.69, 0.24}
\definecolor{grey}{rgb}{0.57, 0.57, 0.57}

\begin{tikzpicture}[scale=0.95, transform shape]
    \def\radii{1/5}; 
    \def\period{1.2};
    \def\rc{0.05*\period}
    
    \coordinate(A0) at (0, 0); 
    \def\ucsz{\period / 2 / cos(30)};
    \def\cylcol{gold}

    \draw[white] ({-4 * \period}, {0}) -- ({3.2 * \period}, {0});
    
    \def\ly{0};
    \def\lyy{0.5 * sqrt(3) * \period};
    \def\cy{(\ly + \lyy) / 2}
    \def\d{0.1}
    
    \draw[-, thick, white](0, {\cy-\period*(1+\d)}) -- (0, {\cy+\period*(1+\d)});
    \draw[-, thick, white]({-\period/2}, \ly) -- ({2*\period}, \ly);
    \draw[-, thick, white]({-\period*0.5}, {\lyy}) -- ({\period*2}, {\lyy});
    \draw[-, thick, white]({-\period/2}, {\cy}) -- ({2*\period}, {\cy});
    
    \draw[->, line width=0.5pt, grey] ({0}, {0}) -- ({2.8*\period}, 0) node[left, xshift=3pt, yshift=-7pt] {$x$};
    \draw[->, line width=0.5pt, grey] (0, 0) -- (0, {{\period * 2.5}}) node[left, , xshift=-1pt, yshift=-3pt] {$y$};
    
    \filldraw [grey] (A0) circle (0.25pt); 
    \foreach \i in {0, 1, 2, 3, 4, 5} {
        \coordinate(this) at ({\period*\radii*cos(\i * 60)}, {\period*\radii*sin(\i * 60)});
        \filldraw [black, fill=\cylcol, line width=0.25pt] (A0){} + (this){} circle (\rc);
    }
    
    \def \layers { {\period}, {2 *  \period}, {2 * \period * cos(30)}}
    \def \rots {0, 0, 1}
    
    \foreach \layer [count=\c,
    parallel foreach=\rot in \rots via \c]
    in \layers {
        \foreach \n in {0, 1, 2, 3, 4, 5} {
            
            \def \centerx {\layer *cos(\n * 60 + \rot * 30)}
            \def \centery {\layer *sin(\n * 60 + \rot * 30)}
            \coordinate(center) at ({\centerx}, {\centery});
            
            \foreach \i in {0, 1, 2, 3, 4, 5} {
                \coordinate(this) at ({\period*\radii*cos(\i * 60)}, {\period*\radii*sin(\i * 60)});
                \filldraw [black, fill=\cylcol, line width=0.25pt] (center){} + (this){} circle (\rc);
                \coordinate(this) at 
                ({\centerx + \ucsz*cos(30+(\i)*60)}, {\centery + \ucsz*sin(30+(\i)*60)});
                \coordinate(prev) at 
                ({\centerx + \ucsz*cos(30+(\i-1)*60)}, {\centery + \ucsz*sin(30+(\i-1)*60)});;
                \draw[solid, line width=0.25pt, grey](prev) -- (this);
            }
        }
    }

    \def \smallhex {2.5 * \period}
    \def \bighex {3.5 * \period * cos(30)}
    \def\resonatorsizes {\smallhex, \bighex};
    \def \checks {1, 2}
    
    \def \xShiftSize {3.2 * \period}
    
    \draw[-, line width=0.25pt, gray]({0}, {\bighex * cos(30)}) -- ({-\xShiftSize - 0.4 * \period}, {\bighex * cos(30)});
    \draw[-, line width=0.25pt, gray]({0}, {-\bighex * cos(30)}) -- ({-\xShiftSize - 0.4 * \period}, {-\bighex * cos(30)});
    
    \draw[<->, line width=0.75pt, black]({-\xShiftSize}, {\bighex * cos(30)}) -- ({-\xShiftSize}, {-\bighex * cos(30)}) node[midway, left, xshift=0pt]{\large $W$};
    
    \foreach \resonatorsz [count=\c,
    parallel foreach=\check in \checks via \c]
    in \resonatorsizes {
        
        \foreach \i in {0, 1, 2, 3, 4, 5} {
            \coordinate(this) at 
            ({\resonatorsz*cos(0+(\i)*60)}, {\resonatorsz*sin(0+(\i)*60)});
            \coordinate(prev) at 
            ({\resonatorsz*cos(0+(\i-1)*60)}, {\resonatorsz*sin(0+(\i-1)*60)});;
            
            \ifthenelse{\check > 1}
            {\draw[solid, line width=1pt, black](prev) -- (this);}
            {\draw[dashed, line width=0.5pt, red](prev) -- (this);}
            
        }
    }

    \filldraw [black, fill=red, line width=1pt] ({0}, {-\period / 2 / cos(30)}) circle (0.12 * \period) node[right, red, xshift=2pt, yshift=6pt]{\large port};
    \filldraw [black, fill=black, line width=0.25pt] ({0}, {-\period / 2 / cos(30)}) circle (0.05 * \period);
    
    \def \xshift {-0.5 * \period * cos(60)}
    \def \yshift {0.5 * \period * sin(60)}
    \coordinate(D0) at ({\smallhex * cos(30) * cos(30) + \xshift}, {\smallhex * cos(30) * sin(30) + \yshift});
    \coordinate(D1) at ({\bighex * cos(30) * cos(30) + \xshift}, {\bighex * cos(30) * sin(30) + \yshift});
    
    \draw[<->, line width=0.75pt, black](D0) -- (D1) node[midway, below, xshift=5pt]{\large $\Delta$};

\end{tikzpicture}
		}
	\end{minipage}
	\caption{
	Hexagonal wire medium cavity. Gray lines indicate the unit cell of the wire medium, and the red dotted line approximates the boundary of the effective wire medium. The parameter $\Delta$ denotes the distance between the wire medium boundary and the cavity walls.
	} \label{fig:resonator} 
\end{figure}

\begin{figure}[h]
	\begin{minipage}{1\linewidth}
		\center{
			\includegraphics[width=1\textwidth]{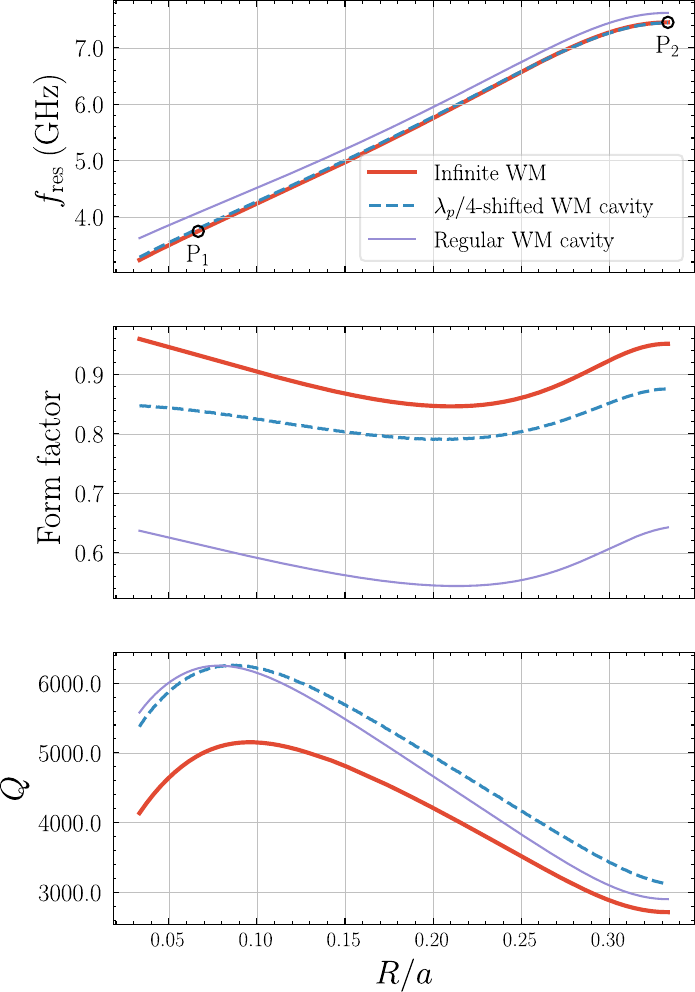}
		}
            \put(-247, 342){\large {(a)}}
            \put(-247, 227){\large {(b)}}
            \put(-247, 112){\large {(c)}}
	\end{minipage}
	\caption{
		(a) Resonance frequency, (b) form factor, and (c) quality factor of the fundamental TM mode in wire medium (WM) cavities as a function of the geometry parameter \( R \), which varies from \( 2r_0 \) to \( a/2 - r_0 \). Introducing a quarter-wavelength air gap between the wire medium and the cavity walls causes the cavity to behave similarly to an infinite wire medium structure, with minimal impact on the quality factor. Points \( \mathrm{P_1} \) and \( \mathrm{P_2} \) correspond to the \( R/a \) values used in the experimental resonators described in the text.
	} \label{fig:comsol_plasma_cavity} 
\end{figure}

The next step of our study was to implement numerical simulations of a finite resonator to assess whether the observed tunability could be realized experimentally. We simulated a hexagonal cavity filled with the metamaterial, as shown in Fig.~\ref{fig:resonator}. The hexagonal cavity shape was chosen to achieve optimal termination of the wire medium, enclosing nearly an integer number of hexagonal unit cells. A typical configuration for wire medium cavities involves enclosing the wire medium with metallic walls placed at the edges of the peripheral unit cells. This arrangement supports a fundamental TM mode with a uniform longitudinal electric field and a sinusoidal transverse field profile, and its resonance frequency lies above the plasma frequency~\cite{balafendiev2022resonator}. In contrast, an optimized configuration introduces an air gap of quarter-wavelength thickness, $\Delta = \lambda_p/4$, between the peripheral unit cells and the cavity walls. This adjustment results in a wire medium bounded by an effective perfect magnetic conductor (PMC) boundary condition~\cite{enriquez2024uniformfieldmicrowavecavities}, which produces a uniform field distribution for the fundamental TM mode and ensures that its resonance frequency coincides with the plasma frequency~\cite{enriquez2024uniformfieldmicrowavecavities}. In this context, TM modes are defined as those in which the electric field is aligned with the wire axes.

Figure~\ref{fig:comsol_plasma_cavity}(a) shows the resonant frequency of the fundamental TM mode in wire medium cavities with an air gap of $\Delta = 0$ (regular cavity) and $\Delta = \lambda_p/4$ (optimized cavity). The results for the cavities indicate that the tuning percentage remains similar to the infinite wire medium, reaching values above $70\%$ (see Fig.~\ref{fig:comsol_plasma_cavity}(a)). Notably, the optimized cavity resonates exactly at the plasma frequency, providing a direct method for measuring the plasma frequency of the wire medium. It is worth mentioning that while a regular cavity approaches the plasma frequency as the number of cells inside the cavity increases \cite{balafendiev2022resonator}, the optimized cavity reproduces the resonance of the infinite wire medium with only a small number of unit cells, as illustrated in Fig.~\ref{fig:comsol_plasma_cavity}(a).  


Although the primary goal of this work is to demonstrate a tunable plasma frequency in a wire medium, the wire medium cavities used for experimental validation are also relevant to axion haloscope applications.  Simulations indicate that the form factor and quality factor (see Figs.~\ref{fig:comsol_plasma_cavity}(b,c)) of both the regular and optimized cavity designs fall within acceptable ranges for such applications~\cite{Bradley2003}, suggesting that the proposed geometry could serve as a viable platform for future dark matter detection experiments. We obtained these values in COMSOL~\cite{comsol} by modeling the wires and cavity walls with an impedance boundary condition corresponding to copper ($\sigma = 5.7 \times 10^7~\mathrm{S/m}$) when computing the quality factor~\cite{balafendiev2022resonator}, and by using Ref.~\cite[Eq.~(13)]{enriquez2024uniformfieldmicrowavecavities} to calculate the form factor. 

\begin{figure*}[t]
	\begin{minipage}{1\linewidth}
		\center{
			\includegraphics[width=1\textwidth]{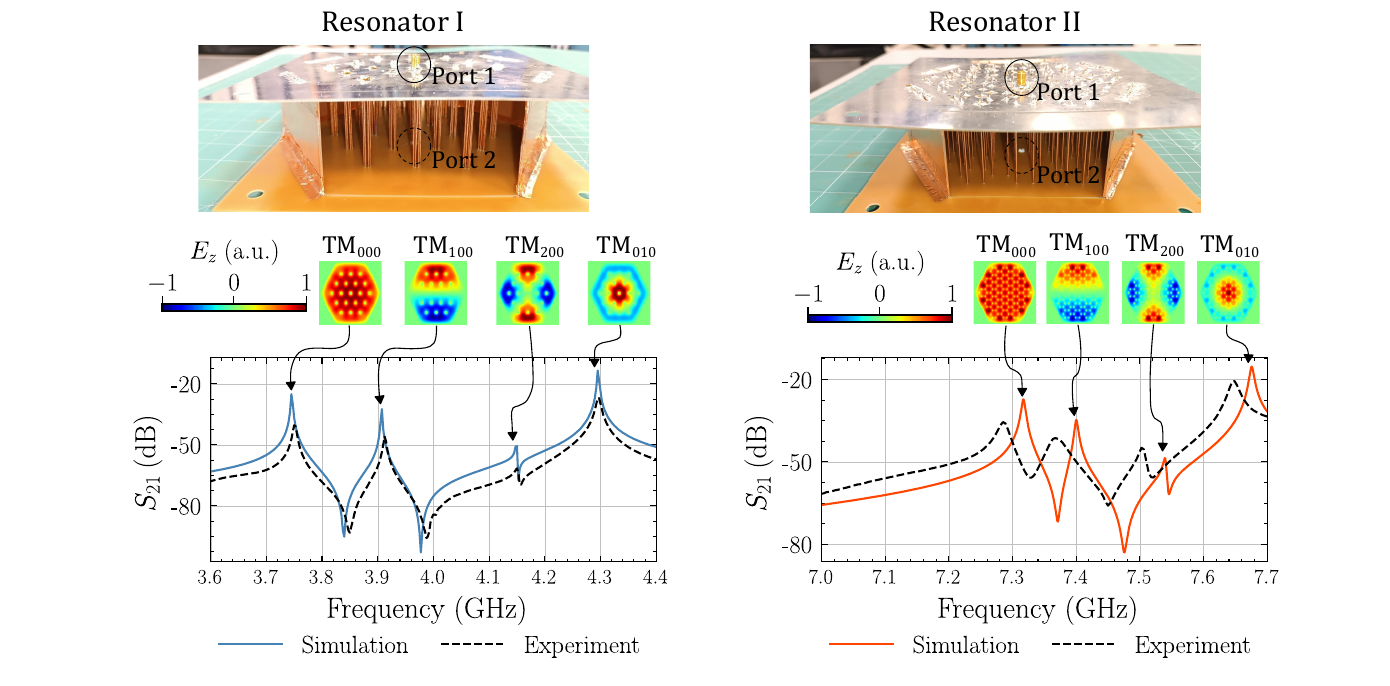}
		}
            \put(-470, 239){\large {(a)}}
            \put(-470, 158){\large {(b)}}
            \put(-470, 112){\large {(c)}}
            \put(-247, 239){\large {(d)}}
            \put(-247, 158){\large {(e)}}
            \put(-247, 112){\large {(f)}}
	\end{minipage}
	\caption{
        Wire medium resonators exhibiting (a--c) low and (d--f) high plasma frequencies. (a,d) Photographs of the experimental resonators; (b,e) simulated field mode profiles; (c,f) simulated and experimental $S_{21}$ parameter. The plasma frequency corresponds to the fundamental resonance frequency. The geometrical parameters of the resonators are listed in Table~\ref{tab:parameters_prototypes}.
	} \label{fig:experiment} 
\end{figure*}
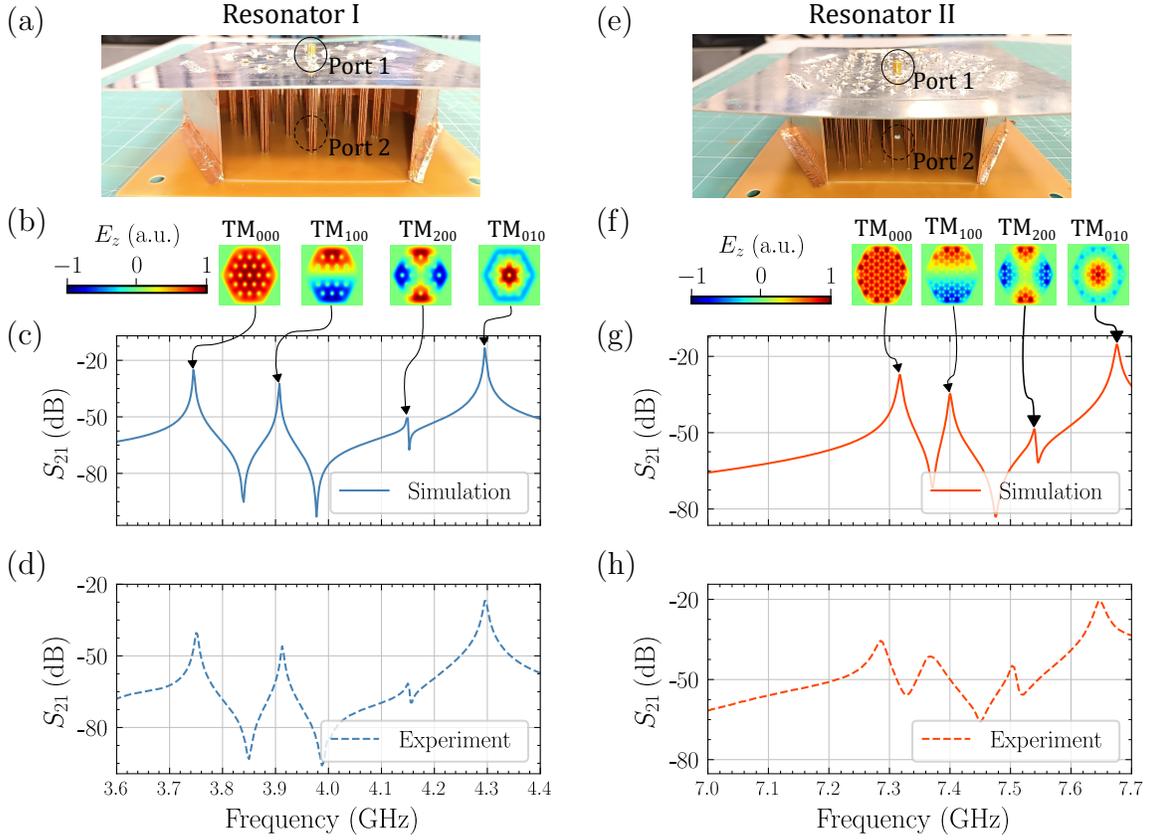



The experimental realization of a mechanically tunable resonator based on the breathing honeycomb structure is beyond the scope of this study, as it primarily poses an engineering challenge. Instead, we conducted a proof-of-concept experiment to demonstrate the feasibility of achieving a wide plasma frequency tuning range using the mechanical deformation method described above.

We fabricated two optimized hexagonal cavities (see Figs.~\ref{fig:experiment}(a,d)). Each cavity was filled with a wire medium based on the breathing honeycomb configuration (see Figs.~\ref{fig:deformation}(a) and~\ref{fig:resonator}), but with different values of the geometric parameter $R$. To ensure that the plasma frequency coincided with the fundamental TM mode resonance of the cavities, we optimized the spacing $\Delta$ between the peripheral unit cells and the cavity walls ($\Delta=\lambda_p/4$). We obtained the plasma wavelength $\lambda_p$ by simulating an infinite wire medium, as shown in Fig.~\ref{fig:deformation} and Table \ref{tab:tunability} (points II and IV). The geometric parameters of the experimental resonators are summarized in Table~\ref{tab:parameters_prototypes}.

\begin{table}[h!]
\centering
    \begin{tabular}{|c|c|c|c|c|c|}
        \hline
        \textbf{Resonator} & $r_0~(\mathrm{mm})$ &$a~(\mathrm{mm})$& $R~(\mathrm{mm})$& $\Delta~(\mathrm{mm})$ & $W~(\mathrm{mm})$ \\ \hline
        I & 0.5 & 30 & 2 & 19.70 & 169.30 \\ 
        II & 0.5  & 30 & 10 & 10.06 & 150.02 \\ \hline
    \end{tabular}
\caption{
    Geometric parameters of the two resonators implemented experimentally. See Figs.~\ref{fig:deformation}(a) and \ref{fig:resonator} for a visual representation. 
}
\label{tab:parameters_prototypes}
\end{table}

In Resonator I, we set $R$ to the minimum value permitted by the constraints of the experimental assembly ($R/a \sim 1/15$). This setup enabled the lowest observable plasma frequency in our measurements. Although this is not the configuration with the smallest possible plasma frequency ($R/a = 1/30$), it allowed us to assign independent holes for each wire, ensuring they remained parallel throughout the structure.

Resonator II used a wire medium with $R/a \sim 1/3$. According to our numerical simulations (see Figs.~\ref{fig:deformation}(c) and~\ref{fig:comsol_plasma_cavity}), this configuration corresponds to the maximum plasma frequency achievable with the proposed tuning method.

We built the experimental resonators using 1 mm-thick double-sided printed circuit boards (PCBs) for the cavity walls. To ensure electrical contact between the wires and the top and bottom plates, we used 1 mm-thick single-sided PCBs with the conductive side facing outward. The top and bottom plates were drilled to accommodate both the wires and the SubMiniature version A (SMA) connector ports. Each resonator consisted of 19 hexagonal unit cells with a lattice constant $a = 3$ cm, enclosed by hexagonal cavity walls. A total of 114 wires with radius $r_0 = 0.5$ mm were soldered into the structures, as shown in Figs.~\ref{fig:experiment}(a,d).

We soldered a 3-mm-long SMA connector to the top and bottom PCBs for signal feeding, as shown in Figs.~\ref{fig:experiment}(a,d). The SMA ports function as monopole antennas, exciting TM modes within the cavities \cite{balafendiev2022resonator}, such that each peak in the $S_{21}$ spectrum corresponds to a resonance mode. We measured the transmission parameter $S_{21}$ using a planar vector network analyzer (VNA, model S5085). Figure~\ref{fig:resonator} shows the exact location of the port in a top view of the wire medium resonators.

To compare with our experimental results, we developed full-wave 3D simulations of the cavities in CST Studio \cite{cst}. The simulations included the exact dimensions and material properties of the experimental resonators. Copper was modeled with a conductivity of $\sigma = 5.7 \times 10^7~\mathrm{S/m}$, while the FR-4 layers of the PCBs were represented as lossy dielectrics with a relative permittivity of $\varepsilon = 4.3$ and a loss tangent of $\tan{\delta} = 0.025$.

Figures~\ref{fig:experiment}(c,f) show the $S_{21}$ parameter for the resonators obtained from simulations and experiments. We focus in particular on the fundamental TM-mode resonance, which corresponds to the plasma frequency in our optimized cavities. This identification is further supported by the uniform field distribution of the $\text{TM}_{000}$ mode observed in the numerical simulations shown in Figs.~\ref{fig:experiment}(b,e). Resonator~I exhibits a plasma frequency of 3.745~GHz in the simulation and 3.750~GHz in the experiment. Resonator~II exhibits a plasma frequency of 7.317~GHz in the simulation and 7.285~GHz in the experiment. The relative error in the plasma frequency between simulations and experiments is approximately 0.1\% for Resonator I and 0.4\% for Resonator II.

Comparing the plasma frequency results for the two resonators, we obtain a tuning percentage of 64.6\% from simulations and 64.1\% from experimental measurements. The ideal tuning percentage predicted for an infinite periodic wire medium, based on our experimental parameters, is 66.1\% (see Table~\ref{tab:tunability}, points II and IV). This estimate, however, assumes perfect material properties and precise dimensions, and therefore does not account for deviations present in the actual experimental resonators.


It is worth mentioning the main sources of discrepancy between our experimental and simulated resonators. On one hand, the excess heat applied during the manual soldering of wires and ports onto the PCBs caused local deformations of the plates, leading to deviations from the ideal behavior assumed in the numerical simulations. In addition, the cavity walls were inserted through the top and bottom plates, requiring a clearance adjustment between the pieces, which introduce variations in the wall spacing $\Delta$. Precise control of this spacing becomes increasingly important at higher frequencies, since the $\lambda_p/4$ distance is smaller. This explains why the discrepancy between the simulation and the experiment in Resonator II appears larger than that in Resonator I. The wall spacing directly influences the resonance frequencies of the TM modes in wire-medium-filled resonators, as discussed in Ref.~\cite{enriquez2024uniformfieldmicrowavecavities}.

In conclusion, we numerically and experimentally studied a honeycomb wire metamaterial composed of hexagonal unit cells and demonstrated the significant tunability enabled by a breathing lattice deformation inspired by topological photonics~\cite{wu2015scheme}. To validate the proposed tuning mechanism, we presented numerical results for both infinite wire medium structures and finite wire medium cavities.

The numerical findings were further validated through full-wave 3D simulations and experimental measurements using two optimized resonators. Resonator I mimics the resonance of an infinite wire medium in a low-frequency configuration, with all six wires positioned closely together near the center of each unit cell. Resonator II corresponds to a high-frequency configuration, in which the wires are spaced at the maximum allowable distance. A comparison of the plasma frequencies in the two resonators revealed a tuning range of 64.1\%.

While previous methods for tuning the operational frequency of plasma haloscopes achieved tuning ranges below 26\%~\cite{kowitt2023tunable, balafendiev2025tunableepsilonnearzero, millar2023alpha}, the present study numerically and experimentally demonstrates the feasibility of exceeding 64\%. Our approach is also competitive with other cavity-based axion search designs: for example, the polygonal coaxial cavity achieves approximately 5\% tuning \cite{DiVora2025}, while the thin-shell axion haloscope reports around 12\% \cite{Dyson2024}. Beyond axion searches, this method constitutes a step forward in structural tunability for metamaterials, where tuning is typically realized by repositioning of the unit cell elements~\cite{lapine_structural_2009, gorkunov_tunability_2008}.

The experimentally observed wide tuning range of the plasma frequency, reaching approximately \( 64\% \), was achieved without altering the unit-cell volume. This result highlights the potential of honeycomb wire media for designing tunable resonators, particularly since both the form factor and quality factor remain adequate for the intended application. Such resonators are well-suited for future dark matter search experiments~\cite{millar2023alpha}, where tuning across a broad frequency range in the GHz regime is essential due to the uncertainty in the axion mass.


\begin{acknowledgments}

The numerical part of this study was carried out with support from the state assignment No. FSER-2024–0041 within the framework of the national project “Science and Universities.” The experimental part was funded by the Russian Science Foundation, grant No. 25-12-00261 (\href{https://rscf.ru/project/25-12-00261/}{https://rscf.ru/project/25-12-00261/}).

\section*{Data availability}
The data that support the findings of this article are
openly available \cite{enriquez_2025_16752795}.
\end{acknowledgments}



\nocite{*}

\providecommand{\noopsort}[1]{}\providecommand{\singleletter}[1]{#1}%

\end{document}